\documentclass[twocolumn,longbibliography,aps,prc,superscriptaddress,showpacs,floatfix]{revtex4-2}
\usepackage{graphics,epsfig,graphicx}
\usepackage{amsbsy,gensymb}
\usepackage{amsmath}
\usepackage{bm}
\usepackage{amsfonts}
\usepackage{cancel}
\usepackage{multirow}
\usepackage{xcolor}
\newcommand {\mcu}{\mathcal{U}}

\begin{document}

\title{The excited state of the $\alpha$-particle: a benchmark study}

\author{P.-Y. Duerinck} 
\affiliation{Physique Nucl\'eaire Th\'eorique et Physique Math\'ematique, C.P. 229, 
Universit\'e libre de Bruxelles (ULB), B-1050 Brussels, Belgium} 
\affiliation{IPHC, CNRS/IN2P3, Universit\'e de Strasbourg, 67037 Strasbourg, France}
\author{A. Deltuva}
\affiliation{Institute of Theoretical Physics and Astronomy,
Vilnius University, Saul\.{e}tekio al. 3, LT-10257 Vilnius, Lithuania}
\author{J. Dohet-Eraly}
\affiliation{Physique Nucl\'eaire Th\'eorique et Physique Math\'ematique, C.P. 229, 
Universit\'e libre de Bruxelles (ULB), B-1050 Brussels, Belgium} 
\author{M. Gattobigio}
\affiliation{
	Universit\'e C\^ote d'Azur, CNRS, Institut  de  Physique  de  Nice,
17 rue Julien Laupr\^etre, 06200 Nice, France}
\author{A. Kievsky} 
\affiliation{Istituto Nazionale di Fisica Nucleare, Largo Pontecorvo 3, 56100 Pisa, Italy}
\author{R. Lazauskas}
\affiliation{IPHC, CNRS/IN2P3, Universit\'e de Strasbourg, 67037 Strasbourg, France}
\author{D. Likandrovas}
\affiliation{Institute of Theoretical Physics and Astronomy,
Vilnius University, Saul\.{e}tekio al. 3, LT-10257 Vilnius, Lithuania}
\author{M. Viviani} 
\affiliation{Istituto Nazionale di Fisica Nucleare, Largo Pontecorvo 3, 56100 Pisa, Italy}

\begin{abstract}
	A benchmark study is performed for the excited state of $^4$He. When the Coulomb interaction is switched off, the $^4$He nucleus exhibits a bound excited state in the vicinity of $p-{}^3$H threshold. As the Coulomb interaction is gradually introduced, the excited state crosses the
	threshold and eventually becomes
	a resonant state. 
    Using three numerical methods, we track the evolution of this excited state and determine the resonance energy and width.  Comparisons of the  theoretical predictions 
    reveal a significant discrepancy with commonly used $R$-matrix values based on the analysis of the experimental data.  We explain the origin for this discrepancy. Additionally, the two-level energy spectrum of $^4$He in the absence of the Coulomb force exhibits characteristics linked to Efimov physics, suggesting a reduced sensitivity to interaction details.
\end{abstract}

\maketitle

\section{Introduction}

The $\alpha$-particle, forming the $^4$He nucleus, is the first double-magic nucleus composed of two protons and two neutrons, with total angular momentum and parity $J^\pi=0^+$. Its binding energy of $28.3\,$MeV represents a binding energy per nucleon of $\sim7\,$MeV -- a remarkably large value compared to the other light nuclei. The $\alpha$-particle clustering provides the bulk of binding energy in medium mass nuclei and holds the key of nuclear binding, since even in the most tightly bound nucleus $^{62}$Ni the binding energy per nucleon of $\sim8.8\,$MeV remains close to the one in $\alpha$-particle. 
Therefore from a theoretical perspective, the description of the
$\alpha$-particle ground state has long been of particular interest. However, due to the complexity of solving the four-body problem, significant advancements in numerical techniques were necessary before accurate results could be obtained using modern two- and three-body nuclear forces~\cite{carlson1998}. 
 A major milestone in this regard was achieved more than two decades ago, when several research groups conducted a benchmark study on the $\alpha$-particle ground state using different numerical methods~\cite{kamada2001}. This study allowed for an assessment of the numerical accuracy achievable by these methods.

Over the years, several computational approaches have emerged as sufficiently flexible to tackle this problem, the
Faddeev-Yakubovsky (FY) equations in configuration and momentum space~\cite{rimas2004,nogga2000}, 
the Alt-Grassberger-Sandhas (AGS) equations~\cite{deltuva:07a}, 
large-basis expansions such as the harmonic oscillator basis~\cite{navratil2000}, the hyperspherical harmonics (HH) method~\cite{kievsky2008}, Gaussian function expansions~\cite{kamimura1989}, and Monte Carlo techniques~\cite{carlson1995}. 
Today, it is widely accepted that the ground state of $^4$He
is well understood in terms of modern two- and three-body nuclear forces. 

Despite its large binding energy per nucleon,  the $\alpha$-particle lacks
stable excited states -- a rare exception among tightly bound nuclei.
However, it exhibits a $0^+$ resonance between
the $p-{}^3$H and $n-{}^3$He thresholds. Evaluation of the experimental data by $R$-matrix 
method locates this resonance at approximately $0.4\,$MeV above the $p-{}^3$H threshold,
with a width of $0.5\,$MeV~\cite{tilley1992}.
To gain deeper insight into the $0^+$ resonance, we can imagine a theoretical experiment in which 
the Coulomb interaction between the two protons in $^4$He  is artificially turned off. If we use 
any of the aforementioned nuclear force models, we immediately observe that the $^4$He nucleus will 
have an excited state just below, but very close to, the triton energy. This suggests that the 
absence of an excited state in the $\alpha$-particle is not due to the specific short-range
properties of the nuclear forces but rather due 
to a repulsive contribution from the Coulomb interaction, which amounts to approximately $0.7\,$MeV 
in $^3$He and $^4$He nuclei. In fact,  when only nuclear forces are considered, this excited state
is so close to the three-body threshold that even a small Coulomb repulsion between the protons is 
enough to push it into the continuum.
This characteristic has been observed before~\cite{fonseca2002,hiyama2004}. It should be noticed
that adding just a isospin-independent repulsive potential, all three- and four-body energies
would move accordingly and the excited $0^+$ state would remain bound.
Thus, to push the excited state to the continuum, the repulsive potential must violate isospin 
symmetry, to keep $^3$H fixed and move $^4$He. Moreover, even in the case in which a short range 
potential would promote the excited state to the continuum, it converts the bound state into a
virtual one, not a resonance. Only when the repulsive potential has a long tail,
the pole from the real negative axis in 2nd sheet moves into complex plane and becomes a resonance
\cite{deltuva:19c}.
Thus, for excited $^4$He to become a resonance one needs isospin violating and a long range force, 
which quite uniquely points to the Coulomb interaction.

In this work, we aim to address two main topics.  First, we perform a numerical benchmark for this resonance state, like the one that has been done for the four-nucleon ground state~\cite{kamada2001} and the one for $p-{}^3$H and $n-{}^3$He scattering above the $n-{}^3$He threshold~\cite{viv2017}. Here two key challenges arise: (1) describing a very shallow state in case the Coulomb interaction is not considered and (2) describing a resonance in case the Coulomb interaction is fully included. Shallow states are hard to describe due to their spatial extension, whereas resonances need a precise description of quickly changing phase shifts, which is computationally challenging. We will show the precision we obtain in our calculations and mention particular challenges, such as the incorporation of various isospin components, $T=0,1,2$, and the role of the $n-^3$He channel.

The second topic we discuss is related to universal physics. The existence of a shallow and a deep four-body bound state attached to a three-body system is a well-known characteristic of bosonic systems close to the unitary limit~\cite{platter2004,hammer2007,kievsky2014,gatto2019}. We will show how universal (Efimov) physics constrains the spectrum of light nuclei, offering a new insight into their structure. Rather than relying on the specific details of nuclear forces  this approach en-globes locations of the deuteron, triton and
$\alpha$-particle inside the universal window~\cite{deltuva2020,kievsky2021}.

The paper is organized as follows: in Section II we briefly describe the
numerical methods used, in Section III we present our numerical results, while
Section IV discusses the properties of the four-nucleon excited state in the context of universal physics. Conclusions are drawn in the last section.

\section{Numerical methods}

The numerical methods used in the present work are the solution of the FY equations
implemented in configuration space, the solution of the operatorial AGS equations
implemented in momentum space, and the HH expansion method. A brief description of the methods is given in Ref.~\cite{viviani2011} and the essential
elements are discussed below.

\subsection{The AGS equations \label{sec:ags}}

The first calculations for scattering processes in the $p-{}^3{\rm H}$, $n-{}^3{\rm He}$, and $d-d$ multichannel system with realistic nuclear potentials and the Coulomb force $V_c(r)$ have been performed in Ref.~\cite{deltuva:07c}. There, the AGS equations  \cite{grassberger:67} for the four-particle transition operators $\mcu^{ji}$ 
were solved in the momentum-space partial-wave representation, whereas the Coulomb force was taken into account via the method of screening and renormalization \cite{taylor:74a,alt:80a,arnas2005}. In short, 
the nuclear two-proton potential is supplemented  by the screened Coulomb one 
\begin{equation} \label{eq:vcr}
V_c(r,R)=  V_c(r)\, e^{-(r/R)^4},
\end{equation} 
where the screening radius $R$ is considerably 
larger than the range of the strong interaction.
Nevertheless, the resulting potential is formally of short range and standard  scattering theory is applicable, just
all the transitions operators acquire the additional parametric dependence
on $R$. For the nucleon-trinucleon scattering the symmetrized transition operators obey the AGS equations
\begin{subequations}
\begin{align}  \nonumber
\mcu^{11}_{R}  = {}& -(G_0 \, t_{R}  G_0)^{-1}  P_{34} 
 - P_{34} \, U^1_{R}\, G_0 \, t_{R} G_0 \, \mcu^{11}_{R} \\
& + {U}^2_{R}   G_0 \, t_{R} G_0 \, \mcu^{21}_{R} , 
\label{eq:U11} \\  \nonumber
\mcu^{21}_{R}  = {}&  (G_0 \, t_{R}  G_0)^{-1} \, (1 - P_{34})
\\ & + (1 - P_{34}) U^1_{R}\, G_0 \, t_{R}  G_0 \,
\mcu^{11}_{R}  \label{eq:tildeU21}
\end{align}
\end{subequations}
where the superscripts 1 and 2 label the 3+1 and 2+2 partitions, respectively. $G_0$ is the free resolvent, $P_{34}$ is the permutation operator,
$t_R$ is the two-nucleon transition operator, while $U_R^1$ and $U_R^2$ are the 3+1 and 2+2 subsystem transition operators; see Ref.~\cite{deltuva:07c} for further details.

The amplitude for elastic $p-{}^3{\rm H}$ scattering with fixed magnitude of the momentum $p$, total angular momentum $J$ and parity $\Pi$ is obtained as the on-shell matrix element
\begin{equation} \label{eq:ptR}
T_R(p,J^\Pi) = 3 \langle \phi_t\,pJ^\Pi | 
\mcu^{11}_{R} |\phi_t\,pJ^\Pi\rangle 
\end{equation} 
with $\phi_t$ being the Faddeev amplitude of the
triton bound state. In the unscreened limit, the long-range part of the amplitude (\ref{eq:ptR}) is given by the 
proton-triton Rutherford amplitude, while the Coulomb-modified short-range part, related to the phase shift in a standard way,
 is 
\begin{gather}      \label{eq:Ttil} 
  \begin{split}
  \tilde{T}(p,J^\Pi) = \lim_{R \to \infty} \left\{ Z_R(p)^{-1} [T_R(p,J^\Pi) - t_R^{pt}(p,J^\Pi)]
     \right\}.
    \end{split}
\end{gather} 
The renormalization factor $Z_R(p)$ and the proton-triton screened Coulomb amplitude $t_R^{pt}(p,J^\Pi)$ are
obtained from the two-body screened Coulomb problem
as explained in Ref.~\cite{deltuva:07c}.

The rate of the convergence with the screening radius $R$ in Eq.~(\ref{eq:Ttil}) depends on the collision energy (or the on-shell momentum $p$), becoming slower with decreasing energy 
\cite{arnas2005,deltuva:07c}. In Ref.~\cite{deltuva:07c} well-converged results were obtained with $R$ ranging from 10 to 20 fm, however, the relative energy between charged clusters was above 1 MeV. In contrast, the present work studies the  $p-{}^3{\rm H}$ scattering below the  $n-{}^3{\rm He}$ threshold,  not  considered in Ref.~\cite{deltuva:07c}. In order to improve the convergence, $R$ values up to 25 fm are used, and the number of partial waves is increased, including the orbital angular momentum $l_x \leq 9$ for two protons, $3N$ partial
waves with spectator orbital angular momentum $l_y \leq 8$ and total angular momentum $J_y \leq \frac{15}{2}$, and $4N$ partial waves
with 1+3 and 2+2 orbital angular momentum $l_z \leq 8$.
Even with these increased cutoffs the convergence in $R$ is not fully achieved below the relative $p-{}^3{\rm H}$ energy of 0.3 MeV, and it is quite poor below 0.1 MeV. However, at those very low energies the convergence is quite monotonic, and the extrapolation based on the two-body model was performed.
In contrast, at higher energies the convergence is oscillatory, as seen also in other systems \cite{arnas2005}.

In calculations of the scattering length without Coulomb (or with small $R$) the theoretical error of few percent is caused by the presence of a nearby pole, leading to slow convergence of the iterative solution method via the double Pade summation \cite{deltuva:07a}.

\subsection{The Hyperspherical Harmonic expansion}

The $p-{}^3{\rm H}$ scattering wave function in a $0^+$ state has asymptotically the relative orbital angular momentum $\ell=0$ and the channel spin $I=0$.
It can be written as a sum of two terms
\begin{equation}
    \Psi_{p-t}^{0^+}=\Psi_C +\Psi_A \ ,
    \label{eq:psica}
\end{equation}

The $\Psi_C$ term describes the system in the region where the four nucleons are close
to each other. It vanishes at large $p-{}^3{\rm H}$ relative distances.
This term is written as a linear expansion $\sum_\mu c_\mu {\cal Y}_\mu(\rho,\Omega)$,
where $\rho$ is the hyperradius and $\Omega$ the hyperspherical variable set. The functions
${\cal Y}_\mu$ form a complete antisymmetric basis constructed in
terms of the HH functions, spin and isospin vectors whereas the hyperradial dependence is given
in terms of Laguerre polynomials (for more details, see Refs.~\cite{kievsky2008,viv2020}). 
In order to improve the convergence properties of the HH basis,
we have included in the expansion basis also a term associated with the closed-channel $n-{}^3$He:
\begin{equation}
  {\cal Y}_C(\rho,\Omega) = {\cal A}\bigg\{
  \Bigl [ Y_{0}(\hat{\bm y}_l)   [ \phi_{h}(ijk)  s_l]_{0}
   \Bigr ]_{0,0}  {e^{-\alpha y_l}\over y_l}(1-e^{-\beta y_l})\bigg\}\ , \label{eq:psioe}
\end{equation}
where $\phi_h(ijk)$ is the ${}^3$He wave function (formed by particles $ijk$), ${\cal A}$ an antisymmetrizer, $s_l$ the spin state of nucleon $l$, $\bm{y}_l$ the $n-{}^3$He relative coordinate, and $\beta$ a variational parameter (for more details see Refs.~\cite{viv2024}).
The asymptotic behavior of this term is governed by $\alpha=\sqrt{2\mu|E-E_h|}$, 
where $\mu$ is the  $n-{}^3$He reduced mass and $E_h$ the ${}^3$He bound state energy. 
The presence of this term helps to describe the $n-{}^3$He clusterization of the system,
which could have a very large spatial extension within the energy range
under study. Configurations of this type are rather difficult to construct using standard HH states ${\cal Y}_\mu$.

The term $\Psi_A$ asymptotically describes the relative
motion of the $p-{}^3{\rm H}$ system. It can be decomposed into a linear
combination of the following functions
\begin{equation}
  \Omega_0^\pm = {\cal A}\biggl\{ [ \phi_t \otimes \phi_p]_{0}
  \left ( f_0(y) {\frac{G_{0}(\eta,q y)}{q y}
          \pm {\rm i} {\frac{F_0(\eta,qy)}{qy}}} \right )\biggr\} \ ,
  \label{eq:psiom}
\end{equation}
where $y$ is the distance between the proton and $^3$H, $q$ is the magnitude of the
relative momentum, $\phi_t$ is the $^3$H wave function, and
$\phi_p$ the spin-isospin proton vector and they are coupled to 0.

In Eq.~(\ref{eq:psiom}), the functions $F_0$ and
$G_{0}$ are Coulomb functions describing the $p-{}^3{\rm H}$ relative radial motion.
The parameter $\eta=\mu e^2/q$, with $e^2\approx 1.44$ MeV fm and
$\mu$ the $p-{}^3{\rm H}$ reduced mass. The function $f_0(y)=1-\exp(-\beta y)$ in Eq.~(\ref{eq:psiom})
has been introduced to regularize $G_0$  at small $y$ values without modifying the
asymptotic behavior, in fact $f_0(y) \rightarrow 1$ for large $y$ values. The specific value of the
non-linear parameter $\beta$ is not important provided that the asymptotic behavior is recovered
in the region in which the strong interaction between the proton and $^3$H is negligible. Therefore,
$\Omega_{0}^+$ ($\Omega_{0}^-$) describes the asymptotic outgoing
(incoming) relative motion and $\Psi_A$ is given by
\begin{equation}
  \Psi_A= \Omega_{0}^- - {\cal S}_{0}(E)\;
     \Omega_{0}^+  \ ,
  \label{eq:psia}
\end{equation}
where ${\cal S}_{0}(E)$ is the $S$-matrix element at the energy $E$. It can be
determined together with the linear coefficients $c_\mu$ occurring in the HH expansion 
from the stationary points of the functional (Kohn variational principle)
\begin{equation}
   [{\cal S}_{0}(E)]= {\cal S}_{0}(E) -{1\over 2i}
        \left \langle\Psi^{0^+}_{p-t} \left |
         H-E \right |
        \Psi^{0^+}_{p-t}\right \rangle
\label{eq:kohn}
\end{equation}
By varying the functional with respect to the unknown parameters, a linear set of equations
is obtained for ${\cal S}_{0}$ and $c_{\mu}$. 
In the case of the $0^+$ bound states, the wave function is expanded in terms
of the HH-spin-isospin antisymmetric vectors
\begin{equation}
	\Psi^{0^+}_{^4{\rm He}}=\sum_\mu d_\mu {\cal Y}_\mu(\rho,\Omega)
\end{equation}
and the linear coefficients $d_\mu$ can be obtained from the Rayleigh-Ritz variational principle
by diagonalizing the Hamiltonian matrix. Both, the linear system and the eigenvalue problem,
can be solved using standard numerical recipes as the Lanczos algorithm.

The expansion of the scattering and bound-state wave functions in terms of the HH basis
is in principle infinite, therefore a truncation scheme is necessary.
The HH functions are essentially characterized by the orbital angular momentum
quantum numbers $\ell_i$, $i=1,2,3$, associated with the three Jacobi
vectors, and the grand angular quantum number $K$ (each HH
function is a polynomial of degree $K$). The basis is
truncated to include states with $\ell_1+\ell_2+\ell_3\le \ell_{\rm
max}$. A value of $\ell_{\rm max}=6$ has been found to be sufficient. 
Between these states, we retain only the HH functions with
$K\le K_{\rm max}$. Moreover, states with total isospin $T=0,1,2$ have been considered.
The largest uncertainty of the method is related to the use of a
finite basis due to the slow convergence of the results with
$K_{\rm max}$.{This problem can be partially overcome by performing
calculations for increasing values of $K_{\rm max}$ and then using some
extrapolation rule to $K\rightarrow\infty$ (see Appendix~B of Ref.~\cite{viv2024} for more details).
The uncertainty introduced by the extrapolation procedure can be used to estimate the
theoretical error due to the truncation of the HH basis.

\subsection{The Faddeev-Yakubovsky equations in configuration space}
Within the framework of Faddeev-Yakubovsky (FY) equations \cite{F61,Y67}, the wavefunction of a four-body system is decomposed into FY components $\mathcal{K}^{l}_{ij,k}$ and $\mathcal{H}_{ij,kl}$ representing all partition chains arising from the subsequent breaking of system $(ijkl)$ into its subclusters. By considering all possible arrangements, it is possible to construct $12$ $\mathcal{K}$-type (3+1) and $6$ $\mathcal{H}$-type (2+2) configurations. Since the FY components constitute natural structures to impose boundary conditions to the wavefunction in asymptotic regions, the FY formalism is particularly suited to treat scattering problems. 

When isospin formalism is employed, likewise solving AGS or HH equations, protons and neutrons are treated as identical fermions (nucleons). The wavefunction can then be constructed from the knowledge of a single $\mathcal{K}=\mathcal{K}^4_{12,3}$ and a single $\mathcal{H}=\mathcal{H}_{12,34}$ component satisfying the following FY equations:
\begin{align}
(E-H_0-V_{12}) \mathcal{K} &= V_{12} \left(P^++P^- \right)\left[(1+Q) \mathcal{K} + \mathcal{H} \right], \label{FYK} \\
(E-H_0-V_{12}) \mathcal{H} &= V_{12} \widetilde{P} \left[(1+Q) \mathcal{K}+\mathcal{H} \right], \label{FYH}
\end{align}
where $H_0$ is the four-body kinetic energy operator, $V_{12}$ is the $NN$ potential between particles $1$ and $2$, and the permutation operators are defined as
\begin{align}
P^+ &= (P^-)^{-1}=P_{23} P_{12}, \\ Q&=-P_{34}, \\ \widetilde{P}&=P_{13} P_{24}.
\end{align}
The four-body wavefunction is then given by
\begin{align}
\Psi &= \left[1+(1+P^++P^-)Q \right](1+P^++P^-) \mathcal{K} \nonumber \\
&+ (1+P^++P^-)(1+\widetilde{P}) \mathcal{H}.
\end{align}
When dealing with short-range potentials, the FY equations \eqref{FYK}-\eqref{FYH} are suited to treat four-body scattering problems in configuration space as the asymptotes of the different binary channels are efficiently separated. If the long-range interactions such as the Coulomb potential are present, the FY components attached to different two-body fragmentation channels do not decouple in the asymptotic region, hindering implementation of boundary conditions. An elegant way to solve this issue has been proposed by Sasakawa and Sawada in Ref. \cite{SS79} for three-body systems. This scheme has later generalized to four-body problems \cite{LC20} and is used in the present work, resulting in modified FY equations. 

The numerical resolution of FY equations is carried out by expanding each FY component in a partial wave expansion.  For a quantum state characterized by a total angular momentum $J$, parity $\Pi$, and isospin projection $m_T$, the FY component $\alpha$ is expressed in its proper set of Jacobi coordinates ($\bm{x}_{\alpha},\bm{y}_{\alpha},\bm{z}_{\alpha}$) as
\begin{equation}
\mathcal{F}_{\alpha}(\bm{x}_{\alpha},\bm{y}_{\alpha},\bm{z}_{\alpha}) = \sum_{n} \frac{F_{\alpha n}(x_{\alpha},y_{\alpha},z_{\alpha})}{x_{\alpha}y_{\alpha}z_{\alpha}} \, \mathcal{Y}^{(F)}_{\alpha n}(\hat{x}_{\alpha},\hat{y}_{\alpha},\hat{z}_{\alpha}), \label{PWE}
\end{equation}
where $F_{\alpha n}$ is the radial function for partial wave $n$ and $\mathcal{Y}_{\alpha n}$ is a generalized spherical harmonics including the coupling of individual angular momenta and given by
\footnotesize
\begin{equation}
\mathcal{Y}^{(K)}_n = \left[\left[\left[l_x (s_1 s_2)_{s_x} \right]_{j_x} \left[l_y s_3\right]_{j_y}\right]_{j_{xy}} \!\!\!\! (l_z s_4)_{j_z}\right]_{J^{\Pi}}
\!\!\!\!\!\!\! \otimes  
\left[\left[(t_1 t_2)_{t_x} t_3\right]_{T_3} t_4 \right]_{T m_T},
\end{equation}
\normalsize
for $\mathcal{K}$ components and
\footnotesize
\begin{equation}
\mathcal{Y}^{(H)}_n= \left[\left[\left[l_x (s_1 s_2)_{s_x} \right]_{j_x} \left[l_y (s_3 s_4)_{s_y} \right]_{j_y} \right]_{j_{xy}} \!\!\!\! l_z \right]_{J^{\Pi}} 
\!\!\!\!\!\!\! \otimes \left[(t_1 t_2)_{t_x} (t_3 t_4)_{t_y} \right]_{T m_T},
\end{equation}
\normalsize
for $\mathcal{H}$ components. The radial functions are expressed as the sum of two terms:
\begin{equation}
F_{\alpha n}(x_{\alpha},y_{\alpha},z_{\alpha}) = F^{(C)}_{\alpha n}(x_{\alpha},y_{\alpha},z_{\alpha})+F^{(A)}_{\alpha n}(x_{\alpha},y_{\alpha},z_{\alpha}),
\end{equation}
where $F^{(C)}_{\alpha n}$ is a core function describing the region of space in which the particles are close to each other, while $F^{(A)}_{\alpha n}$ incorporates the asymptote of the corresponding binary channel. For $3+1$ scattering problems, the asymptotic function is constructed as in Eq. \eqref{eq:psiom} from the three-nucleon wavefunctions and includes, as a parameter, the $S$-matrix to be computed. The core function is here expanded over a three-dimensional Lagrange-Laguerre mesh \cite{B15} as
\begin{equation}
F^{(C)}_{\alpha n}(x_{\alpha},y_{\alpha},z_{\alpha}) =\sum_{i_x,i_y,i_z} c_{\alpha n i_x i_y i_z} \hat{f}_{i_x}(x_{\alpha}) \hat{f}_{i_y}(y_{\alpha}) \hat{f}_{i_z}(z_{\alpha}),
\end{equation}
where $\hat{f}_{i}$ is a regularized Lagrange-Laguerre function and where the coefficients $c_{\alpha n i_x i_y i_z}$ are computed by solving the linear systems resulting from the projection of FY onto basis states. The $S$-matrix is finally computed by using integral and Wronskian relations.

In the present calculations, the partial wave expansion \eqref{PWE} include all spin and isospin states and is truncated by imposing $l_x,l_y,l_z \leq 4$, which was found sufficient for convergence. The results have been computed by considering $20-25$ Lagrange-Laguerre functions for each Jacobi coordinate.

\section{Numerical results}
The potential considered in this study is the chiral
nucleon-nucleon (NN) interactions derived at next-to-next-to-next-to-leading order (N3LO) by Entem and
Machleidt~\cite{entem2003,entem2011}. This potential is one of the modern NN
interactions, based on chiral effective field theory (EFT) concepts, and it is
widely used in nuclear structure and reaction studies. In the present
work, we employ the version  with a $500\,$MeV cutoff, which we refer to as N3LO-EM.

To analyze the binding energies of three-nucleon and four-nucleon systems, we introduce the Coulomb interaction
\begin{equation}
	V_c(r_{ij},\lambda_c)=\lambda_c \frac{e^2}{r_{ij}}\frac{(\tau^i_z+1)}{2}\frac{(\tau^j_z+1)}{2},
	\label{eq:cl1}
\end{equation}
where $r_{ij}$ is the distance between the two protons, and $\tau^i_z$ is the $z$ component of the
isospin vector $\bm \tau^i$, acting on nucleon $i$. The
eigenvalue $\tau^i_z$ is $+1$ for the proton and $-1$ for the neutron. To explore the effect of the Coulomb interaction on the nuclear spectrum, we introduce the parameter
$\lambda_c$, which is varied continuously in the range $[0,1]$.
Setting 
 $\lambda_c=0$ removes the Coulomb interaction, while $\lambda_c=1$ fully incorporates it.

\subsection{Ground state energies of  Three- and Four-nucleon systems }
We begin by analyzing the ground state binding energies (B) of $^3{\rm H}$,
$^3{\rm He}$, and $^4{\rm He}$. The results for the N3LO-EM potential
are presented in Table~\ref{tab:tab1}. As expected, the binding of $^3$H energy remains unchanged as $\lambda_c$ varies. 
However, even at $\lambda_c=0$ the binding energies of $^3$H and $^3$He are not exactly identical.
A small charge symmetry breaking of the strong interaction leads to a difference of $B(^3{\rm H})-B(^3{\rm He})$=0.075\,MeV. 
In Table ~\ref{tab:tab1}, the numbers in parentheses represent the maximum numerical differences between the three computational methods used: FY, AGS, and HH. For three-nucleon systems, the results obtained from these methods differ by at most 1 keV. In the case of four-nucleon systems, the numerical uncertainty increases slightly, reaching a maximum difference of 10 keV.

\begin{table}[h]

\begin{tabular}{c|cccc}
             & $B(^3{\rm H})$ &  $B(^3{\rm He})$ & $\Delta$ & $B(^4{\rm He})$ \\
 $\lambda_c$ & [MeV] & [MeV] & [MeV] & [MeV]  \\
	\hline
    0   &  7.854(1) & 7.779(1) & 0.075(1) &  26.150(4)   \\
	0.1 &  7.854(1) & 7.713(1) & 0.141(1) &  26.074(4)  \\
	0.2 &  7.854(1) & 7.648(1) & 0.206(1) &  25.998(4) \\
	0.3 &  7.854(1) & 7.583(1) & 0.271(1) &  25.921(5)  \\
	0.4 &  7.854(1) & 7.518(1) & 0.336(1) &  25.846(5) \\
	0.5 &  7.854(1) & 7.453(1) & 0.401(1) &  25.768(6) \\
	0.6 &  7.854(1) & 7.388(1) & 0.466(1) &  25.696(6) \\
	0.7 &  7.854(1) & 7.323(1) & 0.531(1) &  25.616(6) \\
	0.8 &  7.854(1) & 7.258(1) & 0.596(1) &  25.546(10) \\
	0.9 &  7.854(1) & 7.193(1) & 0.661(1) &  25.464(10)  \\
	1   &  7.854(1) & 7.128(1) & 0.726(1) &  25.396(10)  \\
	\hline
\end{tabular}
	\caption{The ground state binding energies of the $^3$H, $^3$He, their difference $\Delta$, and the ground state binding energy of $^4$He
	 calculated using the N3LO-EM potential, as a function of the variable $\lambda_c$.
Numbers in parenthesis represent numerical uncertainties due to differences given by
	 the three methods used, FY, AGS and HH. \label{tab:tab1} }
\end{table}

\subsection{The $p-{}^3{\rm H}$ Scattering Length and the Excited State $^4{\rm He}^*$}
We present here results for the  $p-{}^3{\rm H}$ scattering length, $a_{p-t}$, and the binding energy of the excited state of $^4$He, $B(^4{\rm He}^*)$, as functions of the parameter $\lambda_c$.
As mentioned in the Introduction, for small values of  $\lambda_c$, $^4$He has
an excited state. In this regime, $a_{p-t}$ is positive and increases as $\lambda_c$
grows, eventually diverging to $+\infty$ at a critical value of $\lambda_c=\lambda_0$. This divergence occurs when the excited state energy coincides with the
$p-{}^3{\rm H}$ threshold. For values of $\lambda_c$ beyond $\lambda_0$, the scattering length jumps 
from $+\infty$ to $-\infty$,  becoming negative as $\lambda_c\rightarrow 1$. 
In this range, $\lambda_0< \lambda_c\le1$, the excited state moves above
the $p-{}^3{\rm H}$ threshold, entering the continuum spectrum and becoming a resonant state.
In the following, we track the evolution of this state as a function of 
$\lambda_c$,  as summarized in 
 Table~\ref{tab:tab2}. The first three columns, labeled FY, AGS, and HH, follow the
trajectory of $B(^4{\rm He}^*)$ as it approaches the $p-{}^3{\rm H}$ threshold, located at $7.854\,$ MeV.
At $\lambda_c=0.25$ the state remains bound, but by $\lambda_c=0.3$, it has already crossed the threshold.
 Computing these very shallow states is numerically challenging and different approaches have been
used. In the case of FY, the binding energy of the excited state has been obtained as the pole 
of the $S$-matrix computed from the low-energy $s$-wave $p-{}^3{\rm H}$ phase shifts, as well as by searching complex energy values at which the $S$-matrix diverges. 
In the case of AGS equations, the off-shell matrix elements of
four-nucleon transition operators below the $p-{}^3{\rm H}$ threshold have been calculated and, from their energy dependence, the pole position corresponding to the binding energy has been located.
In the HH case, the binding energy of the excited state
has been obtained as the second eigenvalue in a variational calculation of the $0^+$ state
in which the wave function has been expanded in terms of the HH basis. All three methods yield consistent results, with differences in the excited-state energy not exceeding 10 keV and comparable to the variations observed in ground-state calculations.
The results for the $^3$H, $^3$He, $^4$He, and $^4$He$^*$ energies are illustrated
in Fig.~\ref{fig:fig1}, where the trajectory of the $^4$He excited state can be clearly observed.
The figure shows an almost linear dependence on $\lambda_c$, a feature that will be utilized in subsequent analysis.

\begin{figure}[h]
	\includegraphics[scale=0.35]{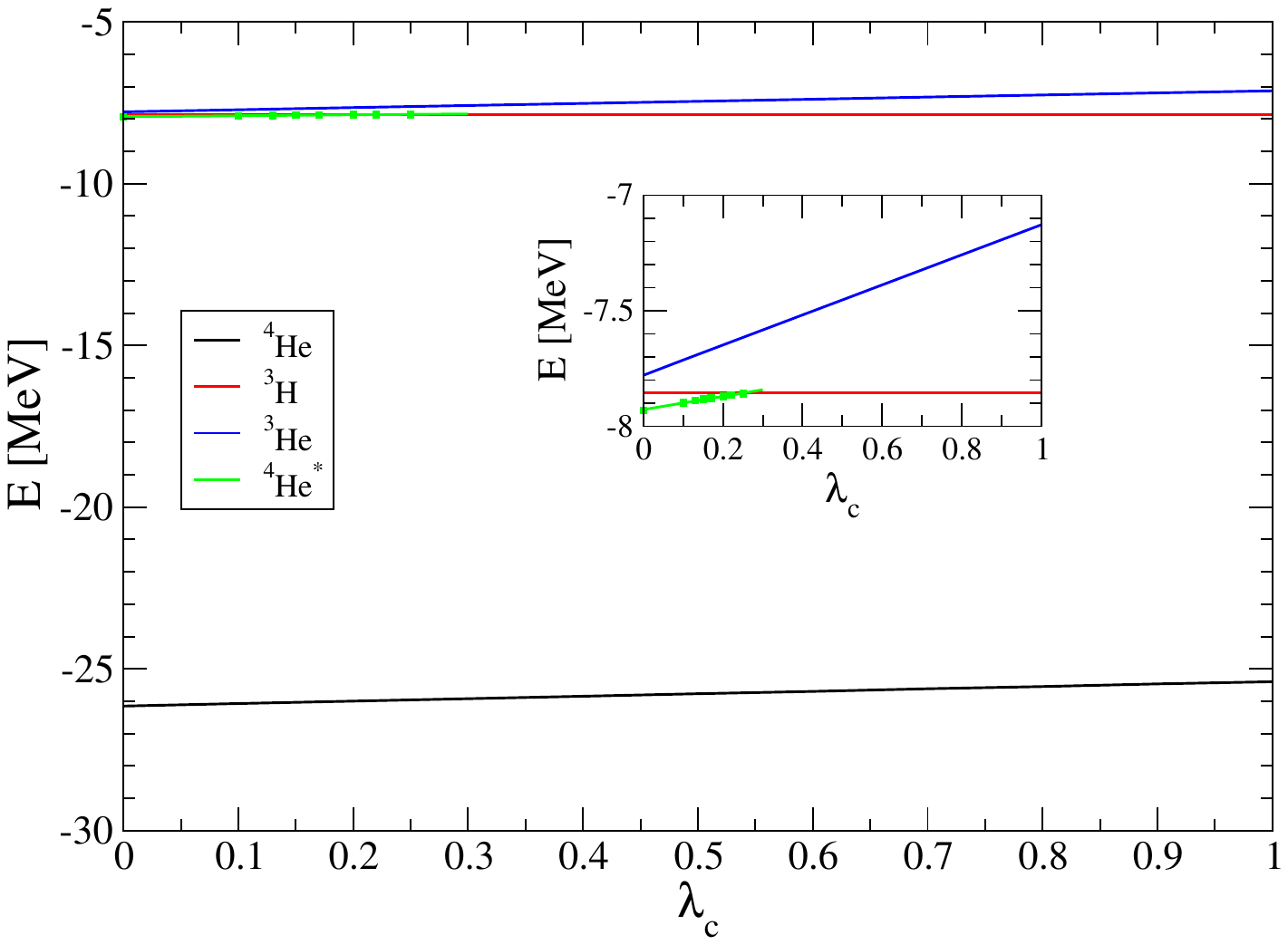}
	\caption{The energies of $^3$H, $^3$He, $^4$He, and
	$^4$He$^*$ as functions of the variable $\lambda_c$. \label{fig:fig1}
}
\end{figure}
As $\lambda_c$ increases, the excited state approaches and eventually crosses the  $p-{}^3{\rm H}$ threshold. To investigate this behavior, we compute the
$p-{}^3{\rm H}$ scattering length, $a_{p-t}$. The second part of Table~\ref{tab:tab2} presents results obtained using the FY and HH methods. The sharp transition in
$a_{p-t}$ values within the range $0.25<\lambda_c<0.3$ further confirms the transition of the excited state into the continuum.
It should be noticed that the HH quoted values are the results
obtained after extrapolating those from the calculations using a very large basis. 
The extended spatial configuration of the excited state is particularly difficult
to describe using the HH basis. Looking at the $a_{p-t}$ results, and considering the theoretical error introduced by 
the extrapolation procedure, the HH results confirm that
the excited state crosses the $p-{}^3{\rm H}$ threshold within the range $0.25<\lambda_c<0.30$.

 In the case of $a_{p-t}$, Table~\ref{tab:tab2} shows the AGS result only for $\lambda_c=0$. In addition, the AGS method follows a different approach for tracing the evolution of the excited state. Instead of scaling the Coulomb potential with a parameter $\lambda_c$ as in Eq.(\ref{eq:cl1}),
a screened Coulomb potential (\ref{eq:vcr})
is introduced and the results are studied in terms
of the screening radius $R$ without performing the renormalization.
For $R=0$ the Coulomb interaction is completely turned off, while in the limit  $R\rightarrow\infty$, the full Coulomb potential would be recovered. The results of this alternative path are presented in Table~\ref{tab:tab3}, showing the binding energy of  $^3$He, binding energy of the $^4$He
excited state  relative to $^3$H,
and the $p-{}^3{\rm H}$ scattering length $a^R_{p-t}$; the latter refers to the screened Coulomb and standard boundary conditions, therefore is different from $a_{p-t}$  in Table~\ref{tab:tab2}.
From the table, one observes that the transition occurs at a screening radius of approximately $R\approx2.6\,$fm. At this point the bound state turns into a virtual state which then evolves into a resonance at larger $R$ with sufficient potential barrier; this kind of transition was observed and explained in Ref.~\cite{deltuva:19c}.

A detailed analysis of the crossing is provided in Fig.\ref{fig:fig2},  where the inverse
of $a_{p-t}$ (upper panel) and $a_{p-t}^R$ (lower panel) are plotted as a function of $\lambda_c$ and $R$ respectively. In the upper panel the FY results have been used and the $\lambda_c$ transition point is identified at $\lambda_0=0.28$   
by interpolation, where the curve crosses the $1/a_{p-t}=0$ axis. In the lower panel the AGS results have been used and the transition point results around $R=2.6\,$fm.
Thus, both approaches -- varying $\lambda_c$ or  $R$ -- indicate the disappearance of the bound excited state of
$^4{\rm He}$ due to inclusion of the Coulomb force. 

\begin{table}[h]
\begin{tabular}{cccc|ccc}
	$\lambda_c$ & \multicolumn{3}{c}{$B(^4{\rm He}^*)-B(^3{\rm H})\,$[MeV]} & 
	\multicolumn{3}{c}{$a_{p-t}\,$[fm]}  \\
	\hline
	    &  FY  &  AGS & HH            &  FY &  AGS & HH  \\
	\hline
	0   & 0.076&0.074 &0.07(1)        & 18.0(2) & 18.7(6)& 23(1) \\ 
	0.1 & 0.044&0.044 &0.03(1)        & 35(2) &     & 38(4) \\ 
	0.13& 0.035&0.035 &               & 43(2) &     &  48(6)    \\ 
	0.15& 0.029&0.030 &               & 52(2) &     & 58(8) \\ 
	0.17& 0.024&0.024 &               & 63(3) &     & 70(10) \\ 
	0.2 & 0.016&0.017 &0.00(1)        & 91(4) &     & 97(20) \\ 
	0.22& 0.011&      &               &       &     & 128(25)     \\ 
	0.25& 0.005&      &               & 266(40) &     & 326(50) \\ 
	0.3 &      &      &               &-363(30) &     &-400(50) \\ 
	1.0 &      &      &               &-18.5(2)  &     &-17(1)  \\ 
	\hline

\end{tabular}
	\caption{The binding energy of the excited state $B(^4{\rm He}^*)$ relative to $B(^3{\rm H})$ and
	  the scattering length $a_{p-t}$ as a function of $\lambda_c$. 
      The theoretical uncertainty for the binding energy is 0.003 MeV for FY and 0.002 MeV for AGS.
      {The numbers in brackets
            in the HH columns are the uncertainties introduced by the extrapolation procedure.} \label{tab:tab2} }
\end{table}

%

\begin{table}[h]
\begin{tabular}{cccc}
        $R\,$[fm] & $B(^3{\rm He})\,$[MeV] &$[B(^4{\rm He}^*)- B(^3{\rm H})]\,$[MeV] & $a^R_{p-t}\,$[fm] \\
        \hline
        0.0 & 7.778 & 0.074  & 18.7         \\
        1.0 & 7.709 & 0.044  & 24        \\
        1.5 & 7.608 & 0.019  & 37         \\
        2.0 & 7.498 & 0.005  & 77    \\
        2.6 & 7.388 &  0     & 2000       \\
        3.5 & 7.280 &        &-71        \\
        4.0 & 7.241 &        &-51        \\
        4.5 & 7.213 &      &  -39        \\
        5.0 & 7.192 &      &  -32        \\
        6.0 & 7.166 &      &  -23        \\
        7.0 & 7.151 &      &  -18.5        \\
        \hline
\end{tabular}
                \caption{$^3$He and the $^4{\rm He}^*$ binding energies and
        the scattering length $a^R_{p-t}$, as functions of the screening radius $R$. The theoretical uncertainties for the binding energy difference and scattering length are 
      0.002 MeV and 3\%, respectively. \label{tab:tab3}}
\end{table}

\begin{figure}[h]
	\includegraphics[scale=0.30]{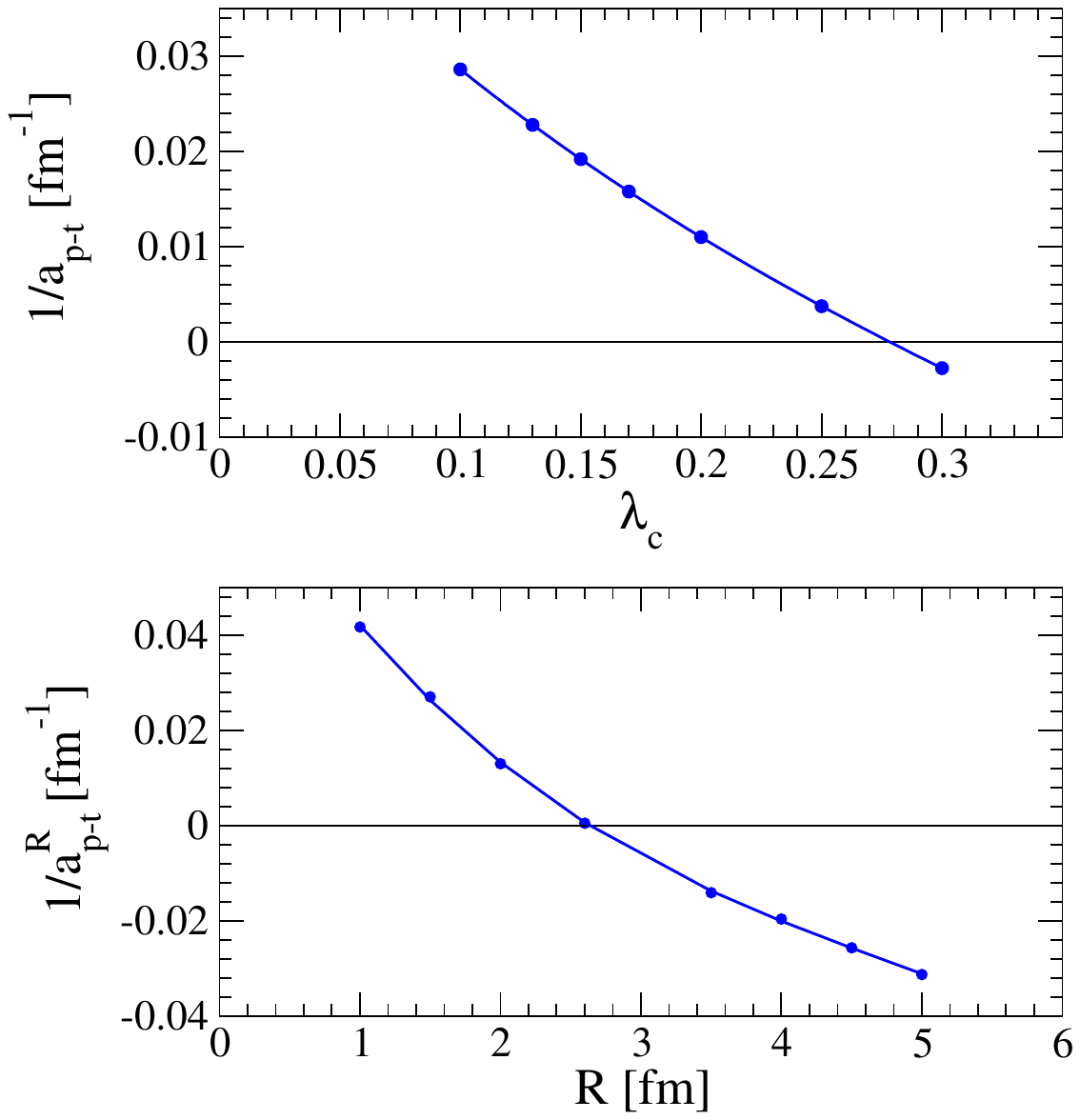}
	\caption{The inverse of the scattering length $a_{p-t}$ as a function of $\lambda_c$
	 (upper pannel) and the inverse of the scattering length $a^R_{p-t}$
     as a function of the screening radius $R$ (lower pannel).   	\label{fig:fig2}}
\end{figure}

\subsection{$p-{}^3{\rm H}$ scattering and the $0^+$ resonance of $^4$He}

In the previous subsection, we examined the evolution of the $^4$He excited state using two approaches: scaling the Coulomb interaction strength and adjusting its range. Here, we focus on the physical scenario where the Coulomb interaction is fully included.
To determine the position and width of the $0^+$ resonance,  we compute the $0^+$ $p-{}^3{\rm H}$ phase shift,
$\delta_0$, at relative energies below $0.72\,$MeV where the $p-{}^3{\rm H}$ is the only open channel. By analyzing the behavior of the phase shift, we can extract the resonance parameters by locating the complex-energy pole of the $S$-matrix. 
To achieve better numerical accuracy, we refine our methods to improve phase shift convergence. In particular, we find it beneficial to explicitly account for the slowly decaying
$n-{}^3{\rm He}$ closed channel asymptote in the scattering wave function. To separate $p-{}^3{\rm H}$  and $n-{}^3{\rm He}$,  the total
isospin $T=0,1$ components should be  considered. In FY and HH calculations, $T=2$ components were also added.
Our results, obtained using the three computational methods, are presented in Table~\ref{tab:tab4} and illustrated in Fig.~\ref{fig:fig3}
as a function of the relative $p-{}^3{\rm H}$ energy. The results in the figure are presented with  method-specific error bars: 
for the FY method, the errors arise from finite basis size and are relatively small;
AGS errors stem from extrapolation accounting for the screened Coulomb interaction employed in the calculations; HH errors originate from extrapolation to an infinite basis. Moreover the $T=2$ component reduces the phase-shift of about $2\%$ at the lowest energy, however the reduction is well below $1\%$ as the energy increases. 

\begin{table}[h]

\begin{tabular}{c|cccc}
	 & \multicolumn{4}{c}{$\delta_0\,$[deg]} \\
        \hline
 $E_{rel}\,$[MeV]  &  FY    & AGS  & &   HH        \\
        \hline
        0.05 &   4.71(8)   & 5.0(20) & &   4.0(1)  \\
        0.1  &   17.9(5)   & 18.0(20) &  &  15.6(5)   \\
        0.15 &   32.9(5)   & 32.9(20)  &      &  28.0(7)    \\
        0.2  &   45.7(5)   & 43.7(20)  &      &  40.1(9)   \\
        0.3  &   62.3(4)   & 61.6(13)   &    &  58.2(16)   \\
        0.4  &   71.5(3)   & 72.2(10)    &    &  66.9(12)   \\
        0.45 &   74.9(3)   & 76.1(9)     &   &  71.7(14)  \\
        0.5  &   77.8(3)   & 79.0(8)     &   &  76.3(15)   \\
        0.6  &   83.5(3)   & 84.3(7)     &   &  81.7(11)   \\
        0.67 &   88.5(2)   & 89.3(6)     &   &  87.9(10)   \\
        0.7  &   92.0(2)   & 92.4(6)     &   &  92.1(10)   \\
        0.72 &   95.9(2)   & 96.2(5)     &   &  94.8(10)   \\
        \hline
	\end{tabular}
       	\caption{ The $0^+$ $p-^3{\rm H}$ phase shift $\delta_0$ as a function
	of the $p-^3{\rm H}$ relative energy $E_{rel}$.
        The AGS results given at 0.05 and 0.1 MeV are obtained by extrapolation (see text).
         The uncertainties in the HH results are introduced by the extrapolation procedure.  \label{tab:tab4}}
\end{table}

\begin{figure}[h]
        \includegraphics[scale=0.35]{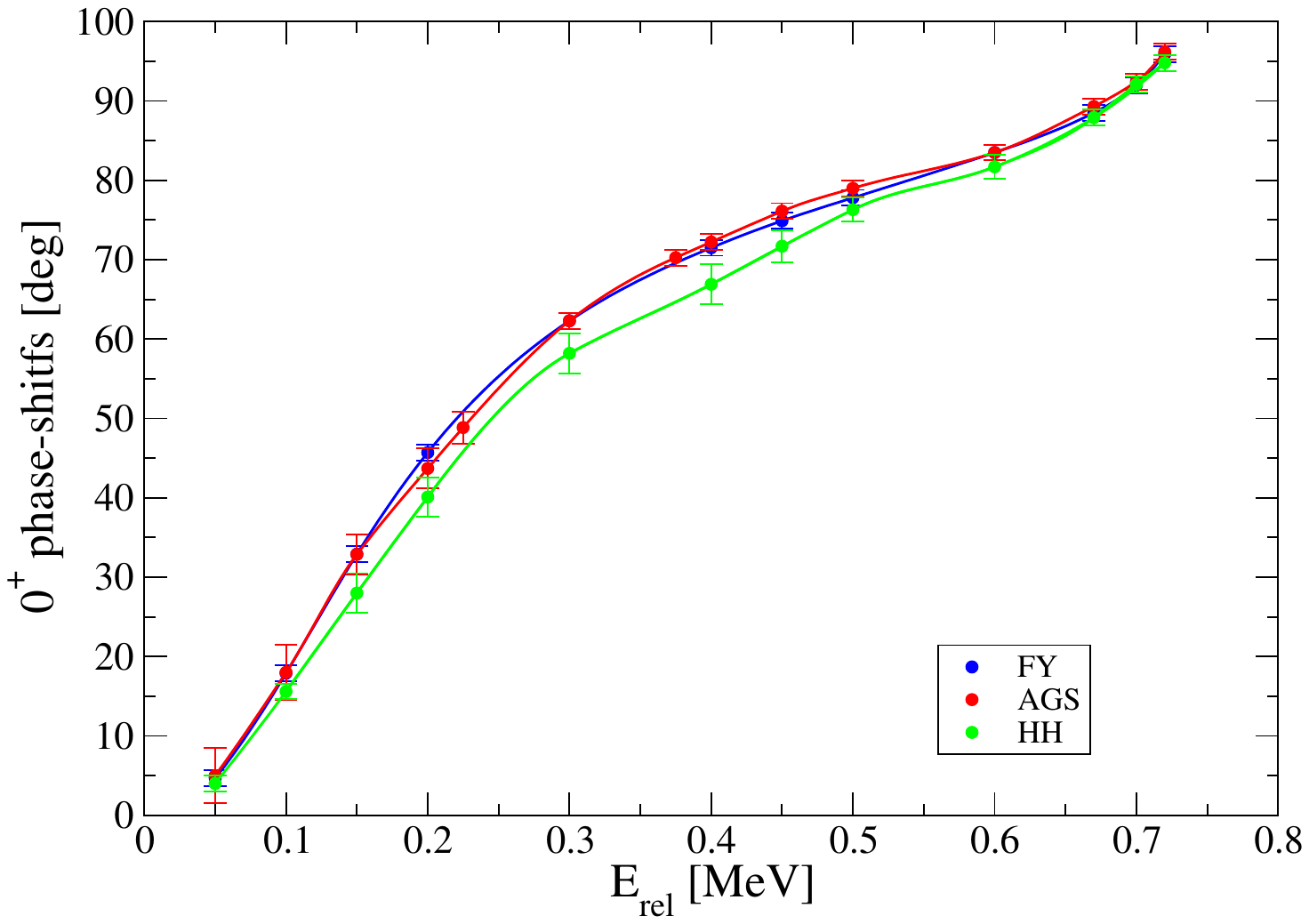}
        	\caption{The $0^+$ phases as a function of the relative $p-{^3\rm H}$ energy. \label{fig:fig3}}
\end{figure}

Using the phase shifts obtained with the FY and HH approaches, we deduce the resonance energy $E_R$ and width $\Gamma$ of the resonance using two different methods.
First,  $E_R$ is determined as the energy at which the derivative
$\delta_0^\prime=\partial \delta_0 / \partial E_R$ reaches its maximum, with the width given by $\Gamma=2/\delta_0^\prime$~\cite{nuc2009}.
The second approach involves locating the $S$-matrix pole. The $S$-matrix is approximated by a rational function whose coefficients are determined from the computed phase shifts~\cite{raki2007}. 
The resonance parameters are then extracted from the complex energy at which the $S$-matrix exhibits a pole. The stability of this pole is assessed by varying the number of computed phase shifts included in the  rational function representing the $S$-matrix~\cite{viv2020}. These two methods yield nearly equivalent results. 

In the AGS calculations the resonance parameters are determined using solutions of the scattering equations above 0.3 MeV, avoiding the lowest-energy region with slow convergence in the screening radius. The  pole of the  transition operator 
is obtained by analyzing the energy dependence of its 
on-shell and  off-shell transition matrix elements, fitting them either with a Laurent series~\cite{deltuva2018} or a continued fraction \cite{schlessinger:68}.
The results from all three methods, including theoretical uncertainties, are presented in Table~\ref{tab:tab5} for the N3LO-EM potential. 
As seen in the table, the three approaches agree well, predicting a resonance energy most likely within the range of 
$0.11\,\,-\,\,0.15\,$ MeV  -- less than half of the value obtained by analyzing experimental data using $R$-matrix method.
 However, there is a reasonable agreement between theoretical and $R$-matrix results for the resonance width, which is approximately $0.5\,$MeV.
It should be noticed that the resonance here is computed using the N3LO-EM potential without including three-body forces.
 Accordingly, the experimental binding energies of the three- and four-nucleon systems are not well reproduced. However, the inclusion of a three-body force moves the threshold to the correct place but substantially does not change the energy of the resonance relative to the $p-{}^3$H threshold~\cite{viv2020}. Moreover, regardless of whether the three-nucleon force is included or not, the $p-^3{\rm H}$ scattering cross sections are well described at low energies. Therefore, the difference between the determined $S$-matrix pole position and the $R$-matrix analysis result might seem surprising. However, this is a well-known phenomenon that occurs for broad resonant states.
 In fact, the $R$-matrix analysis is designed to reproduce experimental data with a minimal set of parameters. As a result, it has natural tendency to introduce an $R$-matrix resonant state centered near the maximum of the scattering cross section at positive energy. If the effective attraction between the projectile and the target is gradually reduced from its critical value, the scattering cross section flattens, while its peak shifts smoothly to higher and higher energies. In this case, the $R$-matrix analysis tends to predict a resonance at increasingly larger positive energies.
 On the other hand, a typical $S$-matrix pole trajectory follows a loop-like shape. At a certain point, as the attraction decreases, the resonance pole undergoes a U-turn and begins moving rapidly into the third energy quadrant. This indicates that the $S$-matrix pole trajectory will inevitably diverge from the one produced by the $R$-matrix analysis based on the same data.
 

\begin{table}[h]

\begin{tabular}{c|cc}
	method   &  $E_R$ [MeV]   & $\Gamma$ [MeV]        \\
        \hline
	FY       &   0.10(1)      & 0.30(10)              \\
	AGS      &   0.15(5)      & 0.50(15)              \\
	HH       &   0.14(6)      & 0.46(15)              \\
        \hline
	Exp. ($R$-matrix)    &   0.39(2)       & 0.50(5)                  \\
        \hline
        \end{tabular}
         \caption{The energy and width of the $0^+$ resonance determined from the
	three methods used to compute the $0^+$ phase shifts        \label{tab:tab5}}
\end{table}

\section{Universal concepts}
As we mentioned in the Introduction, the scheme of two-level bound states, one deep and one shallow,
attached to a three-body state, has been observed before in four-boson systems. The nuclei $^3$H, $^3$He and $^4$He have a dominant symmetrical spatial component so they can be compared to the bosonic case. In fact, the two-level structure observed in $^4$He, non considering the Coulomb force, closely resembles the bosonic case. The spectrum of these light systems is governed by the presence of a shallow bound or virtual state in the two-body system. In the case of the two-nucleon system this refers to the large values of the singlet and triplet scattering lengths. Other ingredients to locate the system inside the universal window (see Ref.~\cite{kievsky2021}) are the
corresponding effective ranges and the energy of the three-body system. With these quantities at hand it is possible to construct a simple two- and three-body potential capable to follow with good accuracy the energy per particle as the number of particles increases. Examples for boson and fermion systems can be found  in Refs.~\cite{kievsky2020,schiavilla2021}, respectively.

Here we would like to use universal concepts to follow the path of the $^4$He excited state as a
function of $\lambda_c$. To this end we follow the method discussed in Ref.~\cite{gatto2023}.
First we construct a two-body Gaussian interaction describing the two-body scattering lengths and effective ranges in the spin $S=0,1$ channels. As a second step we introduce a repulsive hypercentral three-body interaction with range and strength fixed to reproduce the N3LO-EM
ground state energies, $E({}^3{\rm He})$ and $E({}^4{\rm He})$, given in Table~\ref{tab:tab1}, without considering the Coulomb force. For the Gaussian model, the difference $B({}^4{\rm He}^*)-B({}^3{\rm H})$ results $150\,$keV, slightly bigger than the value given by the N3LO-EM potential essentially due to the $s$-wave character of the Gaussian interaction. Then we vary $\lambda_c$ in the range $[0,1]$ and follow the evolution of three- and four-body energies as it is shown in Fig.~\ref{fig:fig4}. As indicated, the figure shows the ${}^3$H, ${}^3$He, ${}^4$He and ${}^4$He$^*$ energies as functions of $\lambda_c$. In the case of ${}^3$He and ${}^4$He the Gaussian interaction predicts slightly less binding whereas in the ${}^4$He$^*$ case slightly more. The N3LO-EM values are given by solid circles. At the critical value $\lambda_0=0.566$ the $\alpha$-particle excited state crosses the $p-{}^3{\rm H}$ threshold and its energy, calculated variationally as the second eigenvalue of the hamiltonian matrix, evolves in the continuum region and at $\lambda_c=1$ it is close to the centroid of the resonance predicted by the N3LO-EM potential. This analysis indicates that the $0^+$ four-nucleon spectrum exhibits a reduced sensitivity to interaction details. This fact has been observed before in the correlation between the triton and $\alpha$-particle ground states called Tjon line~\cite{tjonline} and here we extend this observation to the excited state.

\begin{figure}[h]
        \includegraphics[scale=0.35]{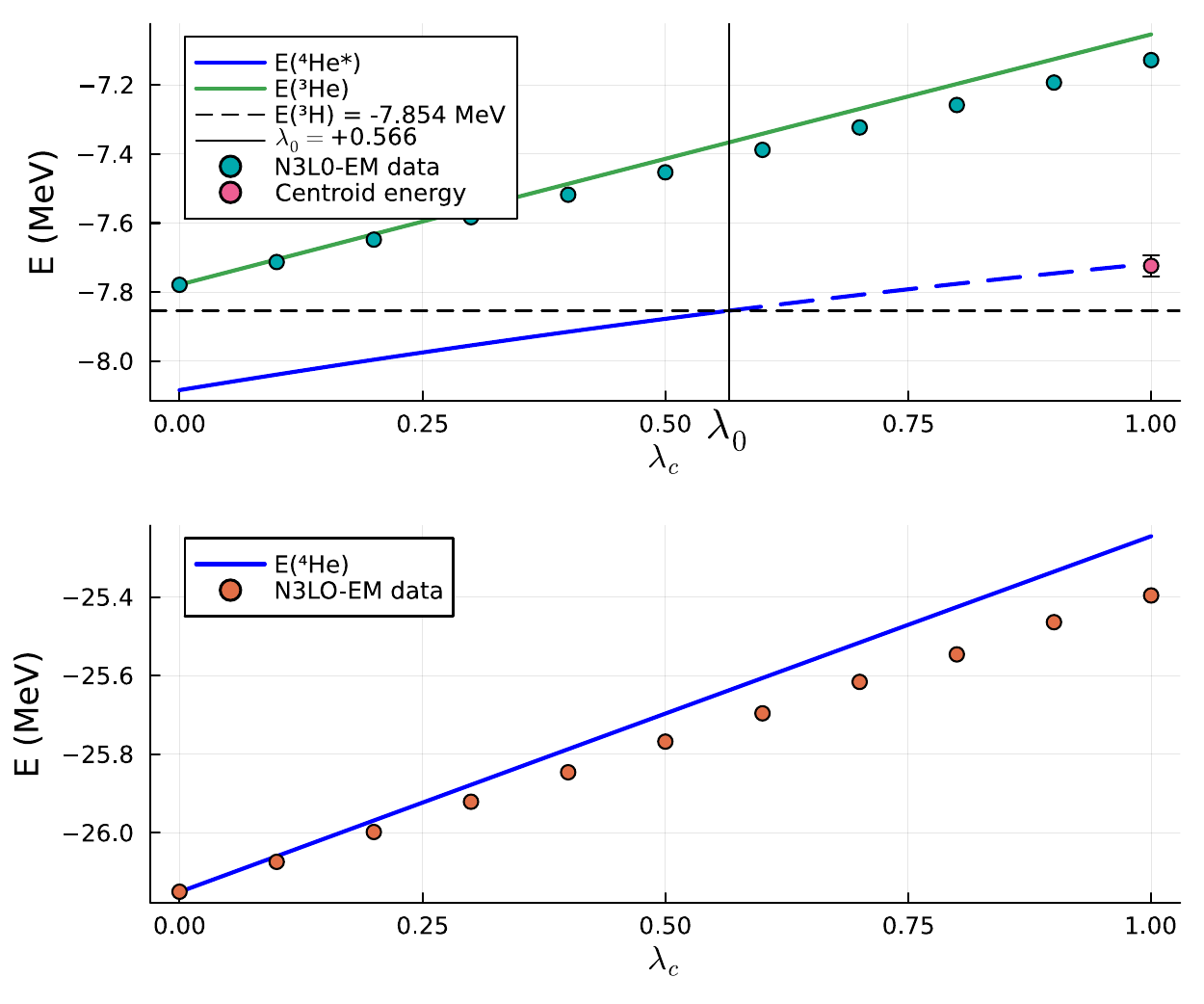}
        	\caption{The  ${}^3$He, ${}^4$He and ${}^4$He$^*$ energies, calculated with a Gaussian potential, as functions of $\lambda_c$ (solid lines). The corresponding N3LO-EM values are given by the solid circles. $\lambda_0$ indicates the value at which the ${}^4$He$^*$ energy crosses the $p-{}^3{\rm H}$ threshold, given by horizontal dashed line, while ${}^4$He$^*$ resonance energy evolution is indicated by the (blue) dashed line. \label{fig:fig4}}
\end{figure}

\section{Conclusions}
In the present work, we have studied the excited state of the $\alpha$-particle, which manifests itself as a resonance just above the $p-{}^3$H threshold. Due to regulatory restrictions on the handling of tritium, no experimental studies have been conducted to determine the precise position of this resonance.
From a theoretical perspective, we have demonstrated that this resonance smoothly connects to a bound excited state when the Coulomb interaction is turned off. At this point, a two-level structure emerges, closely resembling the patterns observed in four-boson systems with large two-body scattering lengths. The universal nature of this state makes it relatively insensitive to the details of the interaction.

Motivated by these considerations, we have conducted a benchmark study of the excited bound state, initially neglecting the Coulomb force. We then gradually introduced the Coulomb interaction using the strength parameter $\lambda_c$ or the range parameter $R$ and followed the evolution of the state. This benchmark study follows a previous analysis of the $\alpha$-particle ground state~\cite{kamada2001} and serves to constrain theoretical uncertainties related to the computational method used to describe the state. Moreover, given the weak dependence of this state on the interaction details, we aim to precisely determine its location relative to the $p-{}^3$H threshold.

In this study, we used the N3LO-EM interaction, a modern chiral-EFT potential that reproduces two-nucleon data with a
 $\chi^2$ per datum close to one. Below the critical
value $\lambda_0$ at which the state crosses the threshold,
the energy is determined with a theoretical uncertainty of approximately 20 keV, an impressive result given its shallow nature. For values of $\lambda_c>\lambda_0$, the
state is no longer bound, which we confirm by computing the $p-{}^3$H scattering length, obtaining negative values as evidence. We then focus on the physical case,  $\lambda_c=1$, where we compute the $0^+$ phase-shifts  for
energies below the $n-{}^3$He threshold, located at $0.72\,$ MeV. From these phase shifts we extract the resonance position and width using two different methods. Our analysis estimates the resonance energy at 
 $E_R\approx 0.13\pm0.05\,$ MeV
with a width $\Gamma\approx 0.5\pm0.2\,$MeV. When including three-nucleon forces, the thresholds shift to their experimental values, though the resonance energy remains nearly unchanged relative to the threshold~\cite{viv2020}. 

It is worth noting that direct comparisons between calculated S-matrix pole positions and resonance parameters obtained from R-matrix analysis of experimental data can be misleading. For broad resonances, there is no straightforward equivalence between these approaches, which explains the discrepancy between our calculated values and those derived from R-matrix analysis of Ref.~\cite{tilley1992}.

We conclude mentioning the universal characteristic of this state, being observed
theoretically in very different systems as for example in the system of four helium
atoms~\cite{hiyama}. Using universal concepts, the existence of this state is not
a consequence of a particular interaction between the components but due to
a discrete scale invariance that constrains the spectrum of the $N$-body
system~\cite{kievsky2014}.
In the case of nucleons, being $1/2$-spin-isospin particles, the dominant
symmetric component stops at $N=4$, so probably the $\alpha$ particle is the
only nucleus where this state can be observed. To show its universal characteristic, we have constructed a simple two- and three-body interaction of
Gaussian form, as discussed in Refs.~\cite{gatto2019,gatto2023},
and used it to trace the state. We have shown that using this simple model for the potential the evolution of three- and four-nucleon systems with $\lambda_c$ follows similar paths as those observed using the N3LO-EM
interaction. We consider this fact a further evidence of the universal behavior in few-nucleon systems.

\vspace{1mm}
\begin{acknowledgments}
This work has received funding from the Research Council of Lithuania (LMTLT) under Contract No.~S-MIP-22-72 and from the Fonds de la Recherche Scientifique - FNRS under Grant No. 4.45.10.08. One of the authors (P.-Y. D.) is a Research Fellow at F.R.S.-FNRS. Computational resources have been provided by the Consortium des Équipements de Calcul Intensif
(CÉCI), funded by the Fonds de la Recherche Scientifique de Belgique (F.R.S.-FNRS) under Grant No. 2.5020.11 and by the Walloon Region. We also have been granted access to the HPC resources of 540 TGCC/IDRIS under the
allocation 2024-AD010506006R3 made by GENCI (Grand Equipement National de Calcul Intensif).
Part of the calculations were made possible by grants of computing
time from the Italian National Supercomputing Center CINECA. We also gratefully
acknowledge the support of the INFN-Pisa computing center.
\end{acknowledgments}

\end{document}